**Potential of Graphene/AlGaN/GaN heterostructures to study the drag and two-stream instability effects**

A. Rehman[1,a)], D.B. But[1,2], P. Sai[1,2], M. Dub[1,2,a)], P. Prystawko[1], A. Krajewska[1,2], G. Cywinski[1,2], W. Knap[1,2], S. Rumyantsev[1]

[1]CENTERA labs, Institute of High Pressure Physics PAS, ul. Sokołowska 29/37, 01-142, Warsaw, Poland.
[2]CENTERA, CEZAMAT, Warsaw University of Technology, ul. Poleczki 19, 02-822 Warsaw, Poland.

a) Author to whom correspondence should be addressed: maksimdub19f94@gmail.com
and  adilrehhman@gmail.com

Graphene/AlGaN/GaN heterostructures are proposed to investigate the drag and two-stream instability effects. In this study, graphene grown by chemical vapor deposition was transferred from copper onto the top of the "standard" AlGaN/GaN wafer, forming a heterostructure with two conducting layers separated by an AlGaN barrier layer. Contacts fabricated to the two-dimensional electron gas and graphene allowed us to study the drag current induced in graphene by passing the drive current through the two-dimensional electron gas. At low temperatures, the graphene drag current exhibited quantum oscillations as a function of the drive voltage. As temperature increases, quantum oscillations disappear, and the magnitude of the drag current increases. Graphene/AlGaN/GaN heterostructures are a promising platform for studying drag and two-stream instability effects, especially if the AlGaN barrier layer thickness can be reduced to a few nanometers.

It is well known that two closely spaced conducting channels interact, leading to the drag effect. The drag effect was first experimentally observed in AlGaAs/GaAs heterostructures[1,2] and later on was extensively studied theoretically and experimentally in graphene/BN/graphene heterostructures[3,4,5]. Apart from the interest to physical mechanisms of the drag effect in different systems, the Coulomb interaction of two conducting layers may have important practical applications due to the possible two-stream instability. This effect is well-known in plasma physics and occurs when different charge carriers or two different beams interact (see Ref.[6] and references therein). In semiconductors, the two-stream instability effect also received significant attention owing to its potential for signal generation at terahertz frequencies. Although the effect was studied theoretically in many publications[7,8,9,10,11,12,13,14], direct experimental observation of two-stream instability in semiconductors is still not reported. We are aware of only one publication where authors observed the low-frequency oscillations at low electric fields in AlGaAs/GaAs single and multiple quantum wells and stated that the two-stream instability might be responsible for such oscillations[15]. However, as the authors also mentioned, the acousto-electric effect and radiative recombination instability may be the real mechanisms behind the observed oscillations.

The semiconductor structures required to observe the drag and two-stream instability effects can consist of two closely spaced quantum wells (QWs) with separate contacts to each QW. Another possibility is to study the system of two graphene layers separated by a thin dielectric, h-BN, for example. Both semiconductor and graphene structures of these types were fabricated and studied in several publications, primarily focusing on the drag effect (see Ref.[4,5] and Refs. therein). One of the problems of the two-stream instabilities experimental observation in semiconductors relates to the difficulty of fabricating the structures that allow achieving high enough drift velocities in both conducting layers with a small leakage current between them.

Additionally, the principal possibility of two-stream instability in double graphene structures is questionable, and it is still being discussed[16,17,18].

In this paper, we propose a new easy-to-fabricate structure consisting of a "standard" AlGaN/GaN wafer for high electron mobility transistors (HEMT) with graphene on top of the barrier layer. The graphene/AlGaN/GaN structures were chosen for two reasons. First, the properties of the charge carriers in these two layers are very different. These are relatively heavy electrons on the AlGaN/GaN interface and light or even massless electrons or holes in graphene. It is known that this situation is very favorable for the two-stream instability development (see, for example, Ref. [9]). Second, previous studies showed that graphene forms a high-quality Schottky barrier on the AlGaN surface[19,20,21,22], providing good isolation between two conducting layers. Therefore, we suppose that graphene/GaN/AlGaN heterostructures are well suited for studying drag and two-stream instability phenomena, which could eventually lead to the generation of radiation at terahertz frequencies[7]. The objective of this work is to investigate the interaction between graphene and two-dimensional electron gas (2DEG) in GaN and to assess the potential of these heterostructures for studying two-stream instability phenomena that could lead to signal generation at terahertz frequencies.

The AlGaN/GaN structures were grown on a sapphire substrate by metalorganic vapor phase epitaxy (MOVPE). Figure 1(a) shows the layers' sequence and their main parameters. The studied devices had the shape of the Hall bars for both AlGaN/GaN heterostructures and graphene (shown in Fig. 1(b)). The Hall bar geometry in the AlGaN/GaN heterostructures was defined by the 150 nm mesas etched using the Inductively Coupled Plasma-Reactive Ion Etching (ICP-RIE) system (Oxford Instruments, Bristol, UK). The single-layer graphene, synthesized via the chemical vapor deposition (CVD) method on copper foils (commercially available from "Graphenea"[23]), was transferred onto the whole AlGaN/GaN heterostructures using an electrochemical delamination process[20]. After that, graphene was patterned in the form of the second conductive layer on top of the AlGaN/GaN Hall bar, as shown in Fig.1 (b) by the red dashed line.

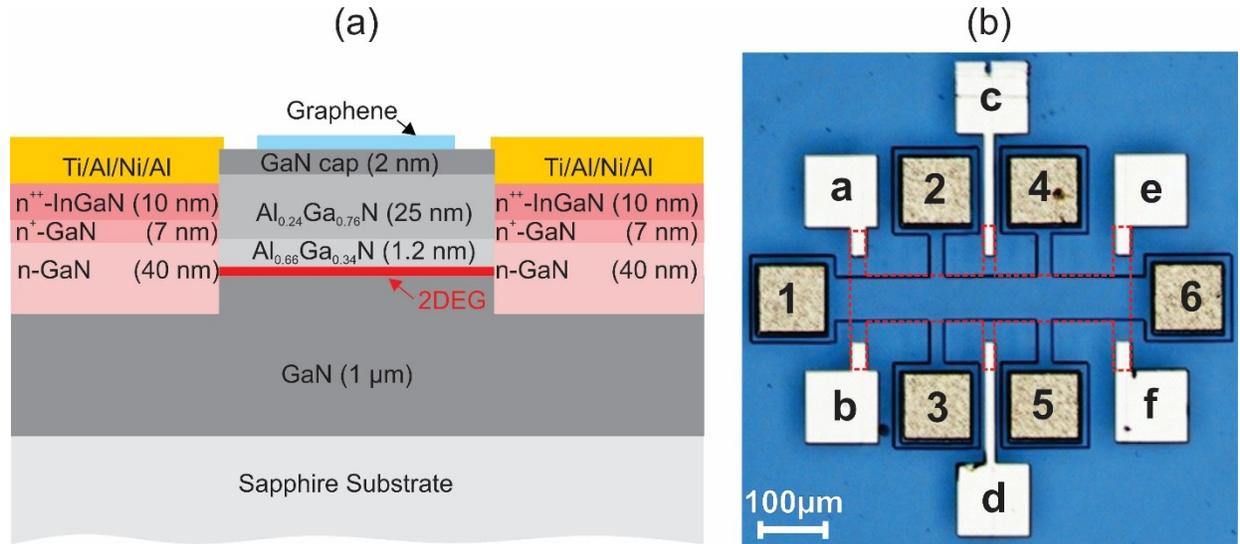

Fig.1 (a) Schematic cross-section of the studied structure along the line between contacts "1" and "6" in panel (b). (b) Optical microscope image of the studied structure, which represents Hall bars for AlGaN/GaN heterostructure and graphene on the top. Graphene is shown by the red dashed line. Pads "1-6" provide contacts to 2DEG, pads "a-f" provide contacts to graphene.

To form low-resistivity ohmic contacts to AlGaN/GaN, we used MOVPE regrowth technique (see Fig.1(a)). The regrown contact regions include 40 nm of n-GaN:Si (n = $1.5 \cdot 10^{19}$ cm$^{-3}$), 7 nm of n$^+$-GaN:Si (n = $5 \cdot 10^{19}$ cm$^{-3}$), and 10 nm of graded n$^{++}$-In$_x$Ga$_{1-x}$N:Si (x = 6-26 % from bottom to top, n = $4.5$-$5 \cdot 10^{19}$ cm$^{-3}$). The details of the regrown ohmic contacts technology can be found in Ref.[24]. The regrowth procedure was followed by thermal deposition of the Ti/Al/Ni/Au (150/1000/400/500 Å) metal stacks on the regrown areas and annealing at 800 °C in a nitrogen atmosphere for 60 s. The described procedure allowed obtaining low contact resistivity $R_c \sim 0.3$ Ohm·mm, which was controlled by linear transmission line model (LTLM) measurements on the test structures fabricated in the same technological process.

The contacts to graphene were formed by thermal deposition of Cr/Au (50/1850 Å) metal stack. All processing was performed using a commercial laser writer system for lithography, based on a 405 nm wavelength GaN laser source (LaserWriter LW405 model, Microtech, Palermo, Italy). The contact pads to 2DEG and graphene are shown in Fig.1 (b) by numbers and letters, respectively.

The studied structures represent two field-effect transistors (FETs): an AlGaN/GaN transistor with the graphene gate and a graphene transistor with the gate formed by 2DEG on the AlGaN/GaN interface. The inset in Fig.2(b) shows these two transistors schematically, with source and drain terminals marked as "S" and "D", respectively. Figure 2(a) shows the output current-voltage characteristics AlGaN/GaN transistor. The electrodes "1" and "6" in Fig.1 (b) served as the source (grounded) and drain of the AlGaN/GaN FET, respectively. Drain current, $I_{dGaN}$, versus drain voltage, $V_{dGaN}$ was measured at different gate voltages $V_{Gr}$ applied to graphene via electrodes "a" and "b", relative to the ground (electrode "1"). Current voltage characteristics in Fig.2(a) represent typical output characteristics of a FET.

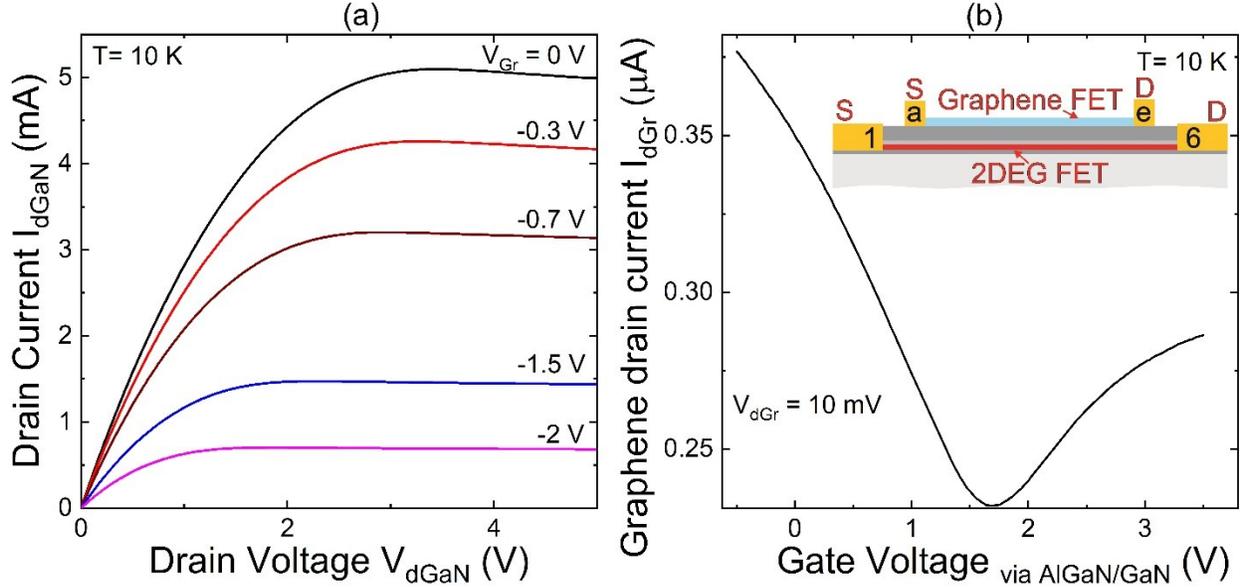

Fig.2. (a) Output current-voltage characteristics of AlGaN/GaN transistor with graphene serving as a gate. (b) Transfer current-voltage characteristics of the graphene transistor with 2DEG serving as the gate. The inset shows schematically graphene and 2DEG interacting transistors.

To characterize the graphene transistor, electrodes "a" and "b" served as the source (grounded), and electrodes "e" and "f" provided the drain contact. The gate voltage, $V_{GaN}$, in this case, was applied to 2DEG via electrode "1". The transfer current-voltage characteristic of graphene transistors is shown in Fig.2 (b). As seen, the charge neutrality voltage for the graphene layer is $V_{cn} \approx 1.7$ V. Therefore, at zero voltage on GaN, graphene is of p-type conductivity.

The Hall measurements at $T = 10$ K yielded the electron mobility and concentration for 2DEG $\mu_{2DEG} \approx 19600$ cm$^2$/Vs and $n = 9 \times 10^{12}$ cm$^{-2}$, respectively. For graphene, the hole mobility and concentration extracted from the Hall measurements at zero voltage on 2DEG were $\mu_{Gr} =$ 1000-1500 cm$^2$/Vs and $p \approx 5 \times 10^{12}$ cm$^{-2}$, respectively.

To measure the drag effect, a voltage was applied to the drain of the AlGaN/GaN transistor (drive voltage $V_{Drive}$), and the current in the graphene layer (drag current $I_{Drag}$) was measured. As seen in Fig.2(a), the current of the AlGaN/GaN transistor (drive current) saturates as a function of the drain voltage (drive voltage) due to the pinch-off at the drain side of the transistor. The maximum drive current depends on the gate voltage and temperature. Close to and above the saturation, the electron concentration in 2DEG decreases from source to drain. The potential between 2DEG and graphene also varies from source to drain. It is zero at the source side, and it is equal to the drain voltage at the drain side. Since 2DEG acts as a gate to graphene, the carrier concentration in graphene also varies from source to drain. The hole concentration decreases towards the drain, and at the drive voltage above the charge neutrality voltage, the type of graphene conductivity changes from holes to electrons. Therefore, at high drive voltage, the drag effect is observed under the condition of a strong non-uniform carrier concentration profile in both conducting layers.

The drag current in graphene was measured by "Keithley Parameter Analyzer series 4200A". This instrument in our setup allowed us to measure reliably currents of the order of 10 pA and above. All measurements were conducted in a closed-cycle cryogenic probe station (Lake Shore Inc., CRX-VF Probe Station).

Figure 3(a) shows the drag current, $I_{Drag}$, as a function of the drive voltage, $V_{Drive}$, at low temperatures. As seen, the drag current fluctuates as a function of the drive voltage. The fluctuations of the drag current and drag resistance as a function of the gate voltage are well known in the system of two graphene layers, manifesting the effect of mesoscopic fluctuations[3,4]. Similar to what was observed before for double-layer graphene structures, these fluctuations are seen up to a relatively high temperature of a few tens of degrees. In Fig. 3(a) these fluctuations are shown as a function of the AlGaN/GaN transistor drain voltage. When the positive drain voltage is applied to AlGaN/GaN transistor, the part of the graphene close to the drain approaches the charge neutrality point, and the fluctuations increase in amplitude as was also observed for double-layer graphene samples (measured as a function of the gate voltage)[3,4]. These fluctuations are well reproducible in shape for different measurements and even, as seen in Fig.3(a), at different temperatures.

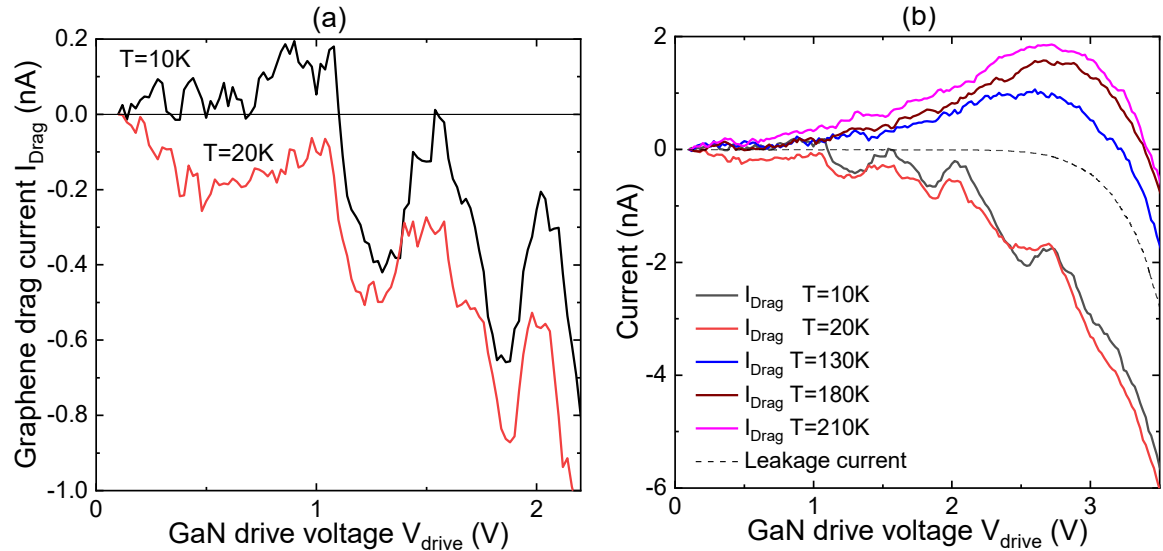

Fig.3. (a) Graphene drag current at two temperatures as a function of the drive voltage applied to the 2DEG (drain of the AlGaN/GaN transistor). (b) The same as in panel (a), but for a wider drive voltage range and at higher temperatures as well. Dashed line shows the leakage current between conducting layers at $T=10$ K.

Figure 3(b) shows the drag current in a wider drag voltage range and at temperatures up to 210 K. It is seen that, with increasing temperature, the drag current changes its sign and increases in amplitude. The dashed line shows the measured leakage current between two conducting layers. It is seen that at drive voltage $V_{\text{Drive}} > 3$ V, the leakage may significantly contribute to the measured drag current. However, at $V_{\text{Drive}} < 2.5$ V, the drag current dominates.

Figure 4 (a) shows the ratio of drag to drive currents at $V_{\text{Drive}} = 1.8$ V as a function of temperature. It is seen that with the temperature increase, the magnitude of this ratio increases, which is consistent with theoretical expectations and experimental results for other systems. Note that due to the temperature dependence of electron mobility in GaN 2DEG, the drive current also depends on temperature. The inset in Fig.4 (a) shows the AlGaN/GaN transistor output characteristics at different temperatures.

The drag effect is often characterized by the drag resistivity, which is defined as:

$$\rho_{Drag} = \frac{V_{Drag}}{I_{Drive}} \frac{W}{L}. \qquad (1)$$

where $W$ and $L$ are the width and the length of the device, respectively. For two interacting conducting layers, the drag resistivity depends on temperature $T$, separation between graphene layers, $d$, and concentrations, $n_1$, $n_2$, in both conducting layers[3,25]:

$$\rho_{Drag} \propto \pm \frac{(k_B T)^2}{d^4} \frac{1}{(n_1 n_2)^{3/2}}, \qquad (2)$$

where $k_B$ is the Boltzmann constant. The sign of $\rho_{Drag}$ depends on the carriers' type, and it is negative when layers have the same type of carriers and positive when layers have different types of carriers[4,3].

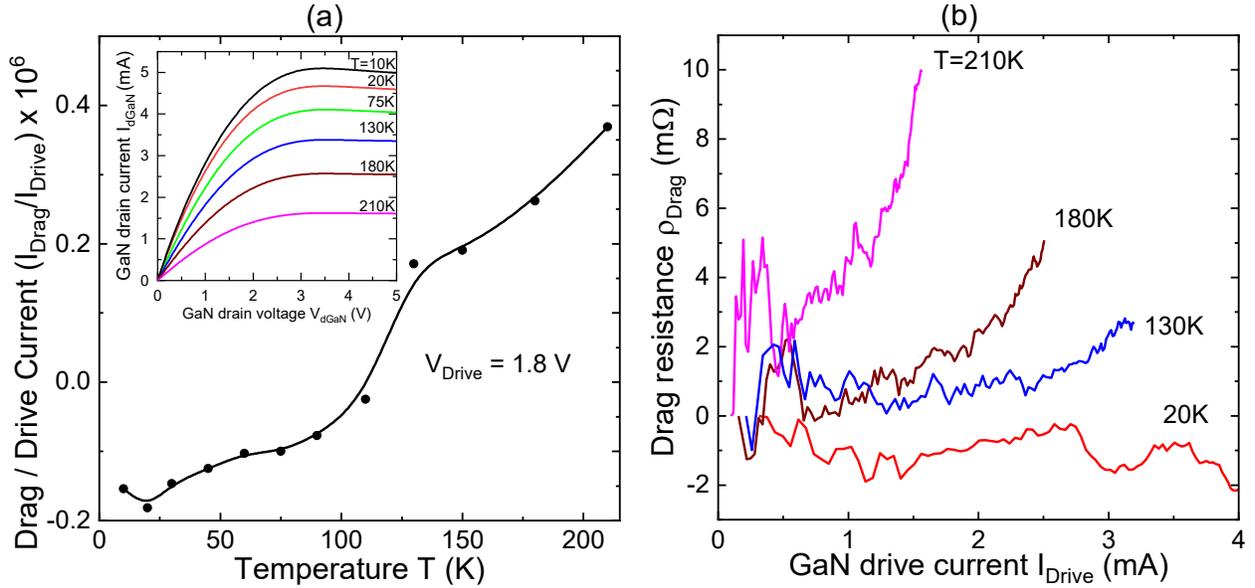

Fig. 4 (a) Ratio of the drag to drive currents as a function of temperature. The inset shows the output current-voltage characteristics of AlGaN/GaN transistor at different temperatures. (b) Drag resistance at different temperatures as a function of the drive current.

Figure 4 (b) shows the drag resistivity versus the drive current for different temperatures calculated as $\rho_{Drag}=I_{Drag}\times R_{Gr}/I_{Drive}\times(W/L)$, where $R_{Gr}$ is the graphene resistance. It is seen that the drag resistance increases with the temperature increase as predicted by Eq. (2).

At temperatures above 130 K, the drag resistivity increases with the increase of the drive current (voltage). It can be explained by the decrease of the electron concentration in the GaN channel and the decrease of the hole concentration in graphene close to the GaN transistor drain when current (voltage) approaches the saturation value (see Eq.(2)).

To conclude, we demonstrated that despite a relatively high separation (~28 nm) between graphene and 2DEG in "standard" AlGaN/GaN heterostructures, the two conducting layers exhibit interaction due to the Coulomb drag. The current drive in 2DEG induces the drag current in the graphene layer, with a magnitude of the order of $10^{-7}$ relative to the 2DEG current. At low temperatures, the quantum oscillations were observed in the graphene drag current, which are well-known for double-layer graphene structures. As temperature increases, the sign of the response changes, and the magnitude of the drag current increases, consistent with prior observations in other systems. Since drag current (drag resistivity) scales with the interlayer separation thickness, $d$, as $1/d^4$, reducing the thickness can enhance the drag effect by more than two orders of magnitude. These findings demonstrate that graphene/AlGaN/GaN heterostructures are a promising platform for studying drag and two-stream instability effects, advancing the understanding of non-equilibrium transport in coupled electron systems, and providing guidance for the design of advanced quantum and high-frequency electronic devices based on engineered heterostructures.

**Acknowledgements**

The work was supported by the European Union through the ERC-ADVANCED grant TERAPLASM (No. 101053716). Views and opinions expressed are, however, those of the author(s) only and do not necessarily reflect those of the European Union or the European Research Council Executive Agency. Neither the European Union nor the granting authority can be held responsible for them. We also acknowledge the support of "Center for Terahertz Research and Applications (CENTERA2)" project (FENG.02.01-IP.05-T004/23) carried out within the "International Research Agendas" program of the Foundation for Polish Science, co-financed by the European Union under European Funds for a Smart Economy Programme. Partial supports were also provided by the National Science Center, Poland, under the research project decision number DEC-2024/08/X/ ST11/01333, by the National Science Centre, Poland grant number 2024/55/B/ST7/02388, and by the ''International Research Agendas'' program of the Foundation for Polish Science co-financed by the European Union under the European Regional Development Fund (No. MAB/2018/9) for CENTERA Labs.

**AUTHOR DECLARATIONS**

Conflict of Interest

The authors declare no conflicts of interest.

**Author Contributions**

A. Rehman: Data curation (lead); Writing– original draft (equal); D.B. But: Writing– review & editing (equal); P. Sai: Methodology (equal); Writing– review & editing (equal); M. Dub: Data curation (equal); Writing– review & editing (equal); P. Prystawko: Methodology (equal); A. Krajewska: Methodology (equal); G. Cywinski: Writing– review & editing (equal); W. Knap: Funding acquisition (lead); S. Rumyantsev: Supervision (lead); Conceptualization (lead); Writing– original draft (equal).

**Data availability**

The data that support the findings of this study will be available on RepOD Repository for Open Data.